\documentstyle[prl,aps,multicol,epsf]{revtex}
\begin{document}
\input epsf
\title{Berry Phase and Spin Qunatum Hall Effect in the Vortex State of
  Superfluid $^3$He in two dimensions}

\author{Jun Goryo{\footnote {Present address; 
Max-Planck Institute for the Physics of Complex Systems, 
Noethnitzer Street 38, 01187 Dresden, Germany}} and Mahito Kohmoto}
\address{\it Institute for Solid State Physics, University of Tokyo,
5-1-5 Kashiwanoha, Kashiwa, Chiba, 277-8581 Japan}

\date{\today}
\maketitle

\begin{abstract}

We show that the spin quantum Hall effect in the 
vortex state of two-dimensional rotating superfluid $^3$He
can be described as an adiabatic spin transport of
Bloch quasiparticles.
We show that the spin Hall conductivity is written
by the Berry phase as well as the Chern number.
The results have similarity to the
adiabatic pumping of Bloch electrons and the spontaneous polarization
in crystalline dielectrics.

\end{abstract}
\draft

\begin{multicols}{2}
\section{Introduction}
The Berry phase (the geometrical phase) arises in quantum mechanical
  systems with an adiabatic change on a closed loop
in a parameter space\cite{Berry}.
In spite of the fact that it is a phase of the wave function,
it could be related to physical effects and, in some cases,
has a connection with topological numbers.
The quantum Hall effect in  Bloch electron systems
is described as an adiabatic charge transport whose process is
closed in a parameter space.
A Berry phase is generated and the quantized Hall conductivity
is written by the Berry phase as well as the Chern number
\cite{TKNN,Kohmoto-85,Simon-Semenoff-Sodano,Kohmoto-93,Goryo-Kohmoto1,Goryo-Kohmoto2}.
Recently, it has been pointed out that  quasiparticles in the vortex
lattice of  $d_{x^2-y^2}$-wave superconductors are
in the Bloch states\cite{FT}.
The spin quantum Hall effect occurs and its conductivity is
written by a Chern number\cite{SQHE-VMT-V}.
Then, one can expect that the spin Hall conductivity
in the vortex state is written by a Berry phase,
when the effect can be described as
an adiabatic spin transport of a closed process.

In this paper, we discuss Bloch quasiparticles
in the vortex state of  $p$-wave superfluid $^3$He in
  two dimensions (2D). We consider a magnetic field with
a weak and homogeneous gradient. Such a field can not be introduced
in  superconductors due to the Meissner effect.
The magnetic field couples to spin through the Zeeman term
and does not to orbital currents because of the neutrality
of the superfluid.
The spin Hall current flows as an adiabatic spin transport.
Its conductivity is written by a Chern
  number and quantized when  an excitation gap exists.
We show that the conductivity is closely related to
the Berry phase.
We also point out that the results have some similarity to the
adiabatic pumping in Bloch electrons\cite{Thouless-pump}
and the spontaneous polarization in
the crystalline dielectrics\cite{King-Smith-and-Vanderbilt-Resta}.
We set $\hbar=c=\mu_{\rm B}=1$, where $\mu_{\rm B}$ is the Bohr
magneton.

\section{Superfluid Helium 3 in two dimensions}

Let $\psi_{\alpha}({\bf x})$ stands for the Fermion field with spin
$\alpha=\uparrow,\downarrow$.
The mean field Hamiltonian for  fermionic superfluid (or
superconductors)
in $D$-dimensional space is
written with the gap matrix $\hat\Delta_{\alpha\beta}({\bf x}, {\bf y})$
as
\begin{eqnarray}
H_{\rm MF}&=&\int d^D x
\psi_{\alpha}^{\dagger} ({\bf x})\epsilon(\hat{\bf p}) \psi_{\alpha}({\bf x})
\\
&&+ \frac{1}{2}
\int d^Dxd^Dy \left[\hat{\Delta}_{\alpha\beta}({\bf x},{\bf y})
\psi_{\alpha}^{\dagger}({\bf x})\psi_{\beta}^{\dagger}({\bf y}) + h.c.\right],
\nonumber\\
&&\epsilon(\hat{\bf p})=\frac{\hat{\bf p}^2 - p_{\rm F}^2}{2 m},
\nonumber
\end{eqnarray}
where $\hat{\bf p}= - i {\bf \nabla}$ and the repeated Greek
indices are summed up.
We may consider the Fourier transform of the gap matrix
in terms of the relative coordinate of ${\bf x}$ and ${\bf y}$,
i.e.
\begin{equation}
\hat{\Delta}_{\alpha\beta}({\bf x}, {\bf y})=\int \frac{d^D p}{(2 \pi)^D}
e^{i ({\bf x} - {\bf y})\cdot {\bf p}}
\hat{\Delta}_{\alpha\beta}({\bf r}, {\bf p}),
\end{equation}
where ${\bf r}=({\bf x} + {\bf y})/2$.

Superfluid $^3$He is in the spin triplet $p$-wave
states\cite{He-3-text}. In general, the gap function for  spin
triplet pairing is 
$\hat{\Delta}({\bf r},{\bf p})
=i \sigma_y {\bf \sigma} \cdot {\bf d}({\bf r}, {\bf p})$,
where ${\bf d}({\bf r},-{\bf p}) = - {\bf d}({\bf r},{\bf p})$ is a
three-dimensional vector in the spin space.
In $p$-wave states, the magnitude of the relative angular momentum
of the Cooper pair $|{\bf l}|=1$ and
${\bf d}$ vectors have a linear dependence on ${\bf p}$.
It is well known that the three phases are observed in the
superfluid $^3$He  (A, B, and A$_1$ phases) \cite{He-3-text}. Those
phases are represented by the different ${\bf d}$ vectors, respectively.

We consider a 2D system. To realize 2D, we introduce
a strong confine potential
along $z$-axis to avoid  quasiparticle excitations along $z$-axis.
The boundary effect introduced by the confine potential locks
the relative angular momentum of all the Cooper pairs in the same direction
  along $z$-axis\cite{He-3-text}. Then, we take $l_z=1$
  ($l_z$: the $z$-component of the angular momentum) in the whole region.
The direction of the ${\bf d}$ vector becomes parallel to the
angular momentum (i.e. ${\bf d} // {\bf e}_z$) because
of the existence of the magnetic dipole interaction which couples spin
  and orbit\cite{He-3-text}.
Then, the ${\bf d}$ vector in our situation is 
\begin{equation}
{\bf d}({\bf r}, {\bf p})={\bf e}_z \phi({\bf r})(p_x + i p_y). 
\label{d-vector}
\end{equation}
This state corresponds to the A-phase\cite{He-3-text}.
Since the direction of the ${\bf d}$ vector and the relative angular
  momentum are frozen, we may neglect the textures and coreless
  vortices\cite{He-3-text}. The Hamiltonian for $^3$He-A is 
\begin{eqnarray}
H_{\rm MF}&=&\int d^2 x
\psi_{\alpha}^{\dagger}({\bf x})
\epsilon(\hat{\bf p}) \psi_{\alpha}({\bf x})
\nonumber\\
&&+
\int d^2x d^2y \left[
\Delta_{\rm A}({\bf x}, {\bf y})
\psi^{\dagger}_{\uparrow} ({\bf x})
\psi^{\dagger}_{\downarrow}({\bf y})
+ h.c.\right],
\label{mf-hamiltonian}\\
&&\Delta_{\rm A}({\bf x}, {\bf y})
=\frac{1}{2}{\rm Tr}[\sigma_x \hat{\Delta}({\bf x},{\bf y})].
\nonumber
\end{eqnarray}

\section{Vortex state in Helium 3 A-phase with a rotation}

It is well known that  rotating superfluid is
direct analogy of  type-II superconductors, and actually the vortex
  states in superfluid $^3$He are detected by the experiments\cite{rotHe-exp}.
Then, we consider  superfluid in a container
that rotates around $z$-axis with an angular velocity $\Omega$.
Hereafter, we use the rotating frame which is fixed on the container.
In the rotating frame, $H_{\rm MF}$ is transformed as
\begin{eqnarray}
H_{\rm MF} \rightarrow H&=&H_{\rm MF} - {\bf \Omega} \cdot {\bf L},
\nonumber\\
&=&\int d^2x \psi_{\alpha}^{\dagger}({\bf x}) \left[
\epsilon(\hat{\bf p} - m {\bf R}) - \frac{m {\bf R}^2}{2}\right]
\psi_{\alpha}({\bf x})
\nonumber\\
&&+ ({\rm pairing ~terms}),
\label{rot-transf}\\
{\bf R}&=&{\bf \Omega} \times {\bf x},
\nonumber
\end{eqnarray}
where ${\bf L}$ is the total angular momentum
of Fermions\cite{Landau-Lifshitz,Anderson-Morel}.
The kinetic energy for the quasiparticle is transformed as
$\epsilon({\bf p}) \rightarrow \epsilon({\bf p} - m {\bf R})
- m {\bf R}^2 / 2$. We consider
$\Omega \sim 1 {\rm rad / s}$ and $|{\bf r}| \leq r_0 \sim 1 {\rm mm}$
  ($r_0$: the radius of the container)\cite{rotHe-exp},
  and we can neglect $- m {\bf R}^2 / 2$ term.
Otherwise, we can cancel out  this term
by introducing a parabolic trap\cite{notes on R^2}. {\it To avoid
this term is essential to introduce the translational invariance we
will discuss below.}

Then, the one to one correspondence can be seen between
our system and  {\it charged} superfluid in a magnetic field
with an infinite London penetration depth, i.e. the strongly type-II
  superconductors. The vector field ${\bf R}$ corresponds to a ``vector
potential ${\bf A}$'' and the Fermion mass $m$
  corresponds to ``the electric charge $e$''. And ``the magnetic field'' is
$2 {\bf \Omega}={\bf \nabla} \times {\bf R}$.

Let us consider a vortex state, i.e. $\Omega > \Omega_{c1}$
and set up a square vortex lattice.
We would like to note that our discussion
is applicable to other types of lattices.
In the vortex state, the gap function has
the singular phase $\varphi({\bf x})$ which satisfies,
\begin{eqnarray}
\Delta_{A}({\bf x},{\bf y})&=&
\tilde{\Delta}_{\rm A}({\bf x}, {\bf y})
e^{-\frac{i}{2}[\varphi({\bf x}) + \varphi({\bf y})]},
\nonumber\\
{\bf \nabla} \times {\bf \nabla} \varphi ({\bf x})&=&
2 \pi {\bf e}_z \sum_i \delta^2({\bf x} - {\bf r}_i),
\label{vorticity}
\end{eqnarray}
where $\tilde{\Delta}_{\rm A}({\bf x}, {\bf y})$ is the
gauge invariant part of the gap function,
${\bf r}_{i}=({\bf e}_x l_i  + {\bf e}_y n_i) a$  with integers 
$l_i$ and $n_i$ is the $i$-th
  lattice point, and ${\bf e}_x$ and ${\bf e}_y$ are the unit vector of
the Cartesian coordinate in the rotating frame.
When $\Omega \sim 1 {\rm rad/s}$ the vortex lattice constant
  $a=\sqrt{\pi / m \Omega} \sim 10^{-2} {\rm cm}$.
 From Eq. (\ref{rot-transf}), the Hamiltonian density operator can be
written in the Nambu representation as
\begin{eqnarray}
&&{\cal {H}}(\hat{\bf p}, {\bf x}, {\bf y})
\label{Hamiltonian}\\
&=&\left(\matrix{
\epsilon(\hat{\bf p} - m {\bf R})\delta({\bf x} - {\bf y}) &
\tilde{\Delta}_{\rm A}({\bf x}, {\bf y})
e^{-\frac{i}{2} [\varphi({\bf x}) + \varphi({\bf y})]}\cr
- \tilde{\Delta}^{*}_{\rm A}({\bf x}, {\bf y})
e^{\frac{i}{2} [\varphi({\bf x}) + \varphi({\bf y})]} & -
\epsilon(\hat{\bf p} + m {\bf R})\delta({\bf x} - {\bf y})}\right).
\nonumber
\end{eqnarray}
The Bogoliubov-de Gennes (BdG) equation\cite{BdG} is
\begin{eqnarray}
\int d^2 y {\cal {H}}(\hat{\bf p},{\bf x},{\bf y}) \Phi_{E}({\bf y})
&=&E \Phi_{E}({\bf x}), 
\nonumber\\
\Phi_{E}({\bf x})&=&(U_E({\bf x}), - V_E^{*}({\bf x}))^{\rm T},  
\label{BdG}
\end{eqnarray}
where, 
\begin{eqnarray}
\psi_{\uparrow}({\bf x})&=&\sum_E 
\left[U_E({\bf x}) \gamma_{E\uparrow} 
+ V_E({\bf x}) \gamma^{\dagger}_{E\downarrow}\right],
\nonumber\\
\psi^{\dagger}_{\downarrow}({\bf x})&=&\sum_E 
\left[-V^{*}_E({\bf x}) \gamma_{E\uparrow} 
+ U^{*}_E({\bf x}) \gamma^{\dagger}_{E\downarrow}\right],
\nonumber
\end{eqnarray}
and $\gamma^{\dagger}_{E\alpha}$ and $\gamma_{E\alpha}$ 
are the creation and annihilation operator of the Bogoliubov quasiparticles,  
respectively. 

Let us discuss the periodicity of the system\cite{Zak-Hofstadter}.
The multivalued 
phase field $\varphi({\bf x})$ which satisfies Eq.(\ref{vorticity}) 
has an ambiguity for deformations which does not change the 
topology of its configuration, i.e. the ambiguity remains  
in terms of the gauge degrees of freedom. So, we may take a constraint  
\begin{equation}
\left\{
\matrix{
\varphi({\bf x} + {\bf e}_x a)&=&\varphi ({\bf x})
- a {\bf e}_x\cdot m{\bf R},
\cr
\varphi({\bf x} + {\bf e}_y a)&=&\varphi ({\bf x})
- a {\bf e}_y\cdot m{\bf R}.}
\right.
\end{equation}
Obviously, it is consistent with Eq.(\ref{vorticity}).
Then, let us define a translation operator
\begin{equation}
T_{\delta {\bf r}}=
\exp\left[i \delta {\bf r} \cdot (\hat{\bf p} + m {\bf R} \tau_3)\right],
\end{equation}
which is the direct analogy of the magnetic translation operator.
The symbol $\tau_3$ denotes the third Pauli matrix in the Nambu
(particle-hole) space. The coordinates are translated by
$T_{\delta {\bf r}}$ as
${\bf x} \rightarrow {\bf x} + \delta{\bf r}$ and
${\bf y} \rightarrow {\bf y} + \delta{\bf r}$.
Then, one can see easily that the operator
$T_{{\bf e}_{x} a}$ and
$T_{{\bf e}_{y} a}$ commute
with ${\cal {H}}({\hat{\bf p}},{\bf x},{\bf y})$, but does not
commute with each other.

We define a unit cell, in which a ``unit flux''
  $2 \pi / m$ penetrates. It is a direct analogy of the
magnetic unit cell, where the magnetic
unit flux $2 \pi / e$ penetrates. A vortex has a ``half unit flux''
  $\pi / m$ and two vortices are contained in a unit cell. Assume that
there are even numbers of vortices, and one may
  choose the unit cell as Fig. 1.
Consider the translations in terms of the cell,
$T_{{\bf e}^{\prime}_x d}=T_{{\bf e}_x a + {\bf e}_y a}$ and
$T_{{\bf e}^{\prime}_y d}=T_{{\bf e}_x a - {\bf e}_y a}$,
where $d=\sqrt{2}a$.
One can see easily that the operators satisfy
\begin{equation}
\left[{\cal {H}}(\hat{\bf p},{\bf x},{\bf y}),
T_{\delta {\bf r}}\right]
=\left[T_{{\bf e}^{\prime}_x d},
T_{{\bf e}^{\prime}_y d}\right]=0.
\end{equation}
Therefore, {\it the eigenstate of ${\cal{H}}(\hat{\bf p},{\bf x},{\bf y})$
are in the Bloch state}, i.e.
\begin{equation}
\Phi_{\bf k}({\bf x})=e^{i {\bf k}\cdot{\bf x}}u_{\bf k}({\bf x}),
\label{Bloch}
\end{equation}
where ${\bf k}$ is in the Brillouin zone (BZ),
$- \pi / d \leq (k_x, k_y) \leq \pi / d$. Here, we omit the band index.
Define 
\begin{equation}
{\cal{H}}_{\bf k}({\bf x},{\bf y}) \equiv 
e^{- i {\bf k}\cdot{\bf x}} 
{\cal {H}}(\hat{\bf p},{\bf x},{\bf y})
e^{i {\bf k}\cdot{\bf y}}.  
\end{equation}
From Eq.(\ref{Hamiltonian}) 
\begin{equation}
{\cal{H}}_{\bf k}({\bf x},{\bf y})
={\cal {H}}(\hat{\bf p}+{\bf k},{\bf x},{\bf y})
e^{- i {\bf k}\cdot({\bf x} - {\bf y})}. 
\end{equation}
Then, from Eq.(\ref{BdG}) and (\ref{Bloch}), 
one can see that the function $u_{\bf k}({\bf x})$ satisfies 
\begin{equation}
\int d^2y {\cal {H}}_{\bf k}({\bf x},{\bf y}) u_{\bf k}({\bf y})
=E_{\bf k}u_{\bf k}({\bf x}),
\end{equation}
and its translation in terms of the unit cell satisfies 
a generalized Bloch condition\cite{Kohmoto-85} 
\begin{eqnarray}
u_{\bf k}({\bf x} + {\bf e}^{\prime}_x d )&=&
\exp\left[i d {\bf e}^{\prime}_x \cdot m {\bf R} \tau_3 \right]
u_{\bf k}({\bf x}),
\nonumber\\
u_{\bf k}({\bf x} + {\bf e}^{\prime}_y d )&=&
\exp\left[i d {\bf e}^{\prime}_y \cdot m {\bf R} \tau_3 \right]
u_{\bf k}({\bf x}).
\label{hpk}
\end{eqnarray}

\vspace{0.5cm}
\begin{figure}
\centerline{
\epsfysize=5cm\epsffile{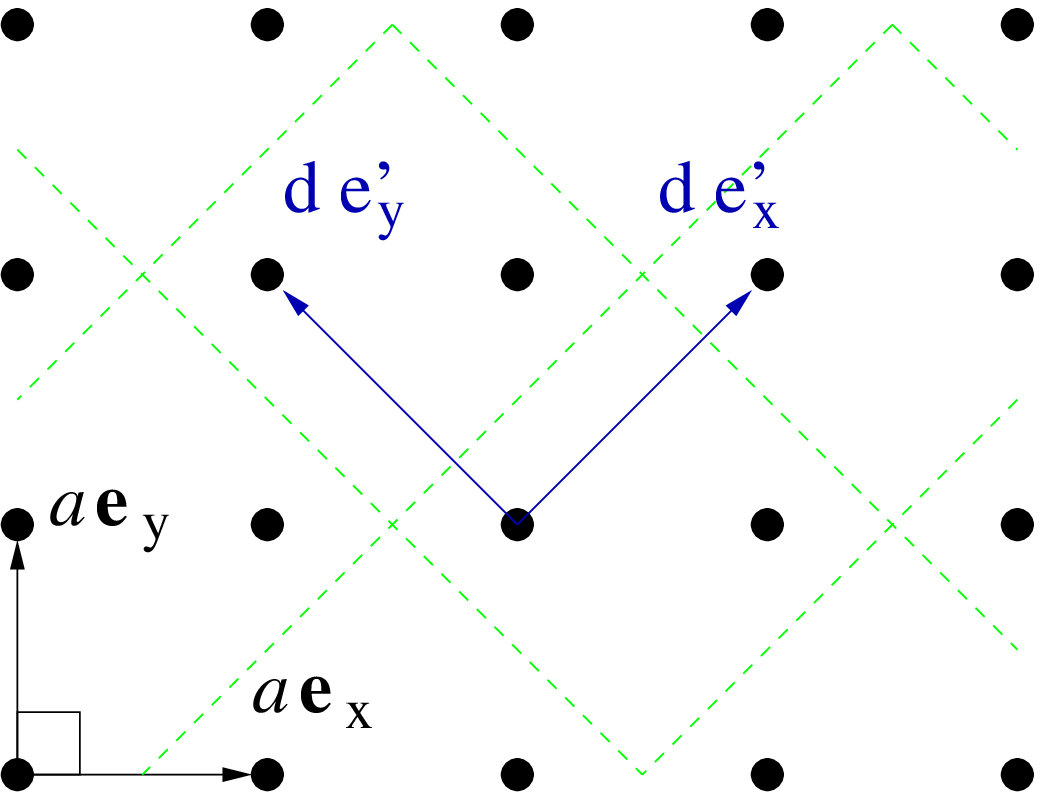}}
\vspace{1cm}
FIG.~1. The unit cells surrounded by green dotted lines.
Black dots denote the vortices.

\end{figure}

The spectrum for  lattice quasiparticles in the square vortex array of
the $p_x + i p_y$-wave superconductors has been investigated
by using the singular gauge transformation\cite{FT,FT2}.
A zero energy state and  gaps around it are found.
In our situation,  the continuum system is considered.
There might be some problems to apply the singular gauge transformation
approach to the continuum systems.
It was pointed out that the quasiparticle spectrum depends on the
choice of the singular gauge transformations\cite{Marinelli-Halperin}.
Here, we assume the existence of an excitation gap. 

\section{The adiabatic process and the Berry phase}

Let us introduce a magnetic field, which is
directed to $z$-axis and has a homogeneous gradient in the rotating
frame, which is written $B_z({\bf x})={\bf x}\cdot\nabla B_z$ and
${\bf \nabla}B_z$ is a constant vector. 
The field will be a driving force
of the spin transport. For a
moment, we consider in the Lagrange formalism. In  superfluid,
the magnetic field couples to spin through Zeeman term and
does not couple to orbital currents.
Then, the Lagrangian is written in the form
\begin{eqnarray}
{\cal L}
&=&\int d^2x \Psi^{\dagger}({\bf x}) \left\{i \partial / \partial_0
- ({\bf x}\cdot{\bf \nabla}B_z /2)\right\} \Psi({\bf x})
\nonumber\\
&&- \int d^2x d^2y \Psi^{\dagger}({\bf x})
{\cal H}(\hat{\bf p},{\bf x},{\bf y}) \Psi({\bf y}), 
\nonumber\\
&&\Psi({\bf x})=
(\psi_{\uparrow}({\bf x}),\psi^{\dagger}_{\downarrow}({\bf x}))^{\rm T}. 
\label{Lagrangian}
\end{eqnarray}
We consider a phase transformation of Eq.(\ref{Lagrangian}),
\begin{eqnarray}
\Psi({\bf x}) \rightarrow
\exp[- i t {\bf x}\cdot{\bf \nabla}B_z / 2] \Psi({\bf x}).
\label{t-dep-transf}
\end{eqnarray}
Then, the term $-({\bf x}\cdot{\bf \nabla} B_z / 2)$ is
absorbed  and the Hamiltonian density operator
is transformed as
\begin{equation}
{\cal H}(\hat{\bf p},{\bf x},{\bf y})
\rightarrow {\cal H}(\hat{\bf p} - {\bf f}(t),{\bf x},{\bf y}),
\end{equation}
where,
\begin{equation}
{\bf f}(t)= t {\bf \nabla} B_z / 2.
\label{f}
\end{equation}
By using the analogy of $U(1)$ (electromagnetic) gauge theory,
we may regard ${\bf f}(t)$ as a vector potential
that couples to the spin current,
since it is introduced by the local spin rotation Eq.(\ref{t-dep-transf}).
We assume that ${\bf f}(t)$ changes adiabatically, i.e.,
  $|{\bf \nabla} B_z|<<1$.  For simplicity,
we write
\begin{equation}
{\cal H}(\hat{\bf p} - {\bf f}(t),{\bf x},{\bf y})
\equiv{\cal H}(t,{\bf x},{\bf y}).
\end{equation}
Then, we solve a time-dependent equation of motion with the adiabatic
  parameter ${\bf f}(t)$
\begin{equation}
i\frac{\partial}{\partial t}\Psi(t,{\bf x})=
\int d^2y {\cal H}(t,{\bf x},{\bf y})\Psi(t,{\bf y}).
\label{time-dep-eq}
\end{equation}
We use the adiabatic approximation
and an eigenvalue equation at fixed $t$ is
\begin{equation}
\int d^2y {\cal H}(t,{\bf x},{\bf y})\Phi_{E(t)}(t,{\bf y})
=E(t)\Phi_{E(t)}(t,{\bf x}).
\end{equation}
Obviously, it is equivalent to the BdG equation (\ref{BdG}) at $t=0$. 
The Hamiltonian ${\cal H}(t,{\bf x},{\bf y})$ has a spatial periodicity
as well as
  ${\cal H}(t=0,{\bf x},{\bf y})$ because ${\bf \nabla}B_z$ is
  homogeneous. Then, eigensolutions
are written in the Bloch form, i.e., $\Phi_{\bf k} (t,{\bf x})
= e^{i {\bf k} \cdot {\bf x}} u_{\bf k} (t,{\bf x})$
(See, Eq. (\ref{Bloch})).
The function $u_{\bf k} (t,{\bf x})$ obeys the equation,
\begin{eqnarray}
\int d^2y {\cal H}_{{\bf k}}(t,{\bf x},{\bf y}) u_{\bf k}(t,{\bf y})&
=&E_{\bf k} (t) u_{\bf k}(t,{\bf x}),
\label{Bloch-form}\\
{\cal H}_{\bf k}(t,{\bf x},{\bf y}) &=&
{\cal H}_{{\bf k} - {\bf f}(t)}({\bf x},{\bf y}),
\label{hkt}
\end{eqnarray}
and hence,
\begin{equation}
u_{\bf k}(t,{\bf x})=u_{{\bf k} - {\bf f}(t)}({\bf x}).
\label{ukt-relation}
\end{equation}
The solution of Eq. (\ref{time-dep-eq}) in the adiabatic approximation
is
\begin{eqnarray}
\Psi_{\bf k}(t,{\bf x})&=&\exp\left[i \int_0^t d t^{\prime}
\left(E_{\bf k}(t^{\prime})
+ \gamma_{\bf k}(t^{\prime})\right)\right] \Phi_{\bf k}(t,{\bf x}),
\nonumber\\
\gamma_{\bf k}(t)&=&
i \int_0^t d t^{\prime}
\left<\Phi_{\bf k}(t^{\prime})\left|
\frac{\partial}{\partial t^{\prime}}\right|\Phi_{\bf k}(t^{\prime})\right>
\nonumber\\
&=&i \int_0^t d t^{\prime}
\left<u_{\bf k}(t^{\prime})\left|
\frac{\partial}{\partial t^{\prime}}\right|u_{\bf k}(t^{\prime})\right>.
\end{eqnarray}
The reciprocal lattice vector for the square vortex array is written 
${\bf G}=(l {\bf e}^{\prime}_x   + n {\bf e}^{\prime}_y)(2 \pi / d) $ 
 with integers $l$ and $n$.  
We  note that it is possible to compactify
the Hamiltonian as
${\cal H}_{\bf k} (t, {\bf x},{\bf y})
\sim {\cal H}_{{\bf k}+{\bf G}}(t,{\bf x},{\bf y})$, 
because they give the equivalent eigensolutions.
Also we note that the parameter ${\bf f}(t)$ varies on the Brillouin zone,
(See, Eq. (\ref{hkt})).
Therefore, when ${\bf f}(t)//{\bf G}$ we have  a period $T$ for a 
closed loop on the parameter space\cite{Zak}.
For example,
\begin{equation}
T=4 \pi / ( |{\bf \nabla}B_z| d), 
\label{T}
\end{equation} 
for ${\bf f}(t) // {\bf e}^{\prime}_x, {\bf e}^{\prime}_y$
and the Berry phases are defined as $\int_0^T dt \gamma_{\bf k}(t)$
for each cases\cite{Berry}.
We introduce the Berry connection,
${\bf a}({\bf k})= \left<u_{\bf k}\left|
{\nabla_{\bf k}} \right|u_{\bf k}\right>$, which is a gauge field
defined on the parameter space\cite{Berry}.
By using Eq. (\ref{ukt-relation}), the Berry phases for the
 parameter ${\bf f}(t) // {\bf e}^{\prime}_x,{\bf e}^{\prime}_y$
are written as
\begin{equation}
\Gamma_x(k_y)=i \int_0^{\frac{2 \pi}{d}} dk_x a_x ({\bf k}),
\label{Berry-x}
\end{equation}
and
\begin{equation}
\Gamma_y(k_x)=i \int_0^{\frac{2 \pi}{d}} dk_y a_y ({\bf k}),
\label{Berry-y}
\end{equation}
respectively.

When ${\bf f}(t)//{\bf G}$, we could write down the Berry phase as
\begin{equation}
\Gamma_{\bf f}=i \oint_{C({\bf f})} d {\bf k} \cdot {\bf a}({\bf k}),
\end{equation}
where $C({\bf f})$ is a closed loop on which ${\bf f}(t)$ moves.
In general, the Berry phase depends on $C({\bf f})$\cite{Berry}.

\section{spin quantum Hall effect and the Berry phase}

Let us calculate the spin current.
In $^3$He-A, the system is invariant under
the spin rotation around $z$-axis
$\Phi_{\bf k}(t) \rightarrow e^{i \theta} \Phi_{\bf k}(t)$ and
${\cal H}(t,{\bf x},{\bf y}) \rightarrow e^{i \theta}
{\cal H}(t,{\bf x},{\bf y}) e^{- i \theta}$.
The spin current ${\bf j}^{s}$ is defined by
the spin conservation law\cite{SQHE-VY}, i.e.,
$\dot{\rho}^s+ {\bf \nabla} \cdot {\bf j}^s=0$, where $\rho^s$ is the
spin density
$(2 \pi)^2 \rho^s({\bf x})=(1/2)\sum_{n \leq 0} \int_{\rm BZ}
d^2k \Psi_{n\bf k}^{\dagger}({\bf x}) \Psi_{n\bf k}({\bf x})$
and we introduce the band index
  $n$. The label $0$ denotes the zero energy.
As we mentioned before, we assume an excitation gap, i.e.
there are no partially filled bands. 
Then, the response of the spin current for the
uniform field ${\bf f}(t)$ is
\begin{eqnarray}
\left<{\bf j}^s(t)\right>&=& \frac{1}{2} \sum_{n<0}\int_{\rm BZ}
\frac{d^2k}{(2 \pi)^2}
\left<\Psi_{n\bf k}^{\dagger}(t)\left|
\frac{1}{i}\left[{\bf r}, {\cal H}(t)\right]
\right|\Psi_{n\bf k}(t)\right>
\nonumber\\
&=&
\frac{i}{2} \sum_{n<0}\int_{\rm BZ} \frac{d^2 k}{(2 \pi)^2}
\left[
\left<\dot{u}_{n {\bf k}}(t)\left|\frac{\partial
u_{n{\bf k}}(t)}{{\partial} {\bf k}}\right.\right> - h.c.\right]
\nonumber\\
&=&- \sigma^s_{xy}\left[{\bf \nabla} B_z \times {\bf e}_z\right],
\nonumber\\
\sigma^s_{xy}&=&\frac{1}{8 \pi}\sum_{n<0} N^{(n)}_{\rm Ch}, 
\label{SQHE-1}
\end{eqnarray}
where 
\begin{eqnarray}
N^{(n)}_{\rm Ch}=
\int_{\rm BZ}\frac{d^2k}{2 \pi i}
\left[{\bf \nabla}_{\bf k} \times {\bf a}_n({\bf k})\right]_z,
\label{Chern}
\end{eqnarray}
is the Chern number for the $n$-th band and 
${\bf a}_n({\bf k})$ is equivalent to the Berry connection for 
the $n$-th band. The detailed calculation is written, for example, in 
Ref.\cite{Goryo-Kohmoto2}.

The Chern number takes integer. 
The reason is based on the fact that ${\bf a}_n({\bf k})$ is defined on
the torus (the BZ) and the Chern number becomes finite if and only if 
${\bf a}_{\bf k}$ has a non-trivial topology.
The nature of the Chern number has been discussed
in detail in Ref.\cite{Kohmoto-85}.

Then, Eq. (\ref{SQHE-1}) shows that a stationary  
spin Hall current flows as an adiabatic spin transport 
and its conductivity is quantized as an integer multiple
of $1/8 \pi$. The same result for the conductivity 
in the vortex state of
$d$-wave superconductors have been obtained\cite{SQHE-VMT-V}. 
The discrete conductance change is expected to occur
when one varies $\Omega$ and $p_{\rm F}$.
  
Here, we calculated the expectation value of the total spin current 
directly by using the adiabatic approximation and obtained  
the spin Hall conductivity. One can see the 
fact in the calculations of
Eq. (\ref{SQHE-1}) that this approach is equavalent to 
calculate the Kubo formular for the spin Hall conductivity 
argued in Ref. \cite{SQHE-VMT-V}.

Finally, we show that $\sigma^s_{xy}$ could be written 
in terms of the Berry phase
\cite{Simon-Semenoff-Sodano,Kohmoto-93,Goryo-Kohmoto1,Goryo-Kohmoto2}. 
By using the Stokes'
  theorem, we have the relation (See, Eq.(\ref{Berry-x}),
Eq.(\ref{Berry-y}), Eq. (\ref{SQHE-1}) and Eq. (\ref{Chern})),
\begin{eqnarray}
\sigma^s_{xy}&=&
\frac{-1}{16 \pi^2} \sum_{n<0} \times
\label{spin-Hall-Berry}\\
&&\left[\int_{0}^{\frac{2 \pi}{d}} dk_x  \frac{d \Gamma^n_y(k_x)}{d k_x}
- \int_0^{\frac{2 \pi}{d}} dk_y \frac{d \Gamma^n_x(k_y)}{d k_y}\right].
\nonumber
\end{eqnarray}

\section{Relations to the other arguments}

The effect have some similarity to
the adiabatic pumping which is originally argued by
  Thouless and discussed actively at present\cite{Thouless-pump}.
In pumping, an adiabatic ac perturbation yields a dc current,
and the charge transfer per a cycle is independent of the period
of the perturbation. The charge transfer is quantized when the
ac perturbation is commensurate with the lattice in 1D.
As we mentioned before, the Hamiltonian in our system
${\cal {H}}_{\bf k}(t)$ is compactified
and moved periodically by the adiabatic parameter ${\bf f}(t)$.
Then, the Hamiltonian changes ac-like, and the change
yields a dc spin Hall current.
To make a correspondence to the Thouless' arguments,
one calculates a spin transfer per period $T$.
Assume that ${\bf f}(t)={\bf e}^{\prime}_y t |{\bf \nabla} B_z| / 2$.
The spin Hall current flows along $x^{\prime}$-axis (See, Fig. 1)
and the spin transfer per the boundary of the unit cell along
$y^{\prime}$-axis is
\begin{equation}
\Delta S_z= d \int_0^T dt \left< j^s_{x^{\prime}} (t) \right>
=- \sum_{n < 0}\frac{N^{(n)}_{\rm Ch}}{2}  
\label{spin-transfer}
\end{equation}
(see, Eq. (\ref{T}), (\ref{SQHE-1}) and (\ref{Chern})). 
The result does not depend on $T$ .
It comes from the fact that both of the magnitude of the quantized current
and $T^{-1}$ are proportional to $|{\bf \nabla} B_z|$.
We emphasize that the spin transfer
is quantized, i.e., {\it the integral spin transfer occurs}.
The result is analogous to the Thouless' result\cite{Thouless-pump}.

The value $\Delta S_z / d$
corresponds to the magnetization change per the period.
From Eq. (\ref{spin-Hall-Berry}), the magnetization change
is written by the Berry phase.
Then, the present result is also similar to the spontaneous
polarization of crystalline dielectrics
, which is written by the
Berry phase introduced by a closed adiabatic change of the
Kohn-Sham potential\cite{King-Smith-and-Vanderbilt-Resta}.

Essentially, the similarity comes from
the fact that the effects argued here
are caused by the closed adiabatic change
in the Bloch states with the finite energy gap.
A parallel discussion for the present arguments
have been made in the Bloch electron systems in the presence
of the electromagnetic field with respect to the charge transport
\cite{Goryo-Kohmoto2}.

\section{summary and discussions}

In summary, we consider Bloch quasiparticles in a vortex state of
superfluid $^3$He-A in 2D with a rotation along $z$-axis.
A magnetic field is along $z$-axis with a weak
homogeneous gradient in the rotating frame.
The field could be represented by
an adiabatically changing vector potential which couples to the spin current.
The adiabatic process is defined on a closed loop in the parameter
  space (the Brillouin Zone) and generates a Berry phase.
The spin Hall current flows in the process. 
We calculated the expectation value of the total spin current 
directly by using the adiabatic approximation and obtain the 
spin Hall conductivity. This approach is 
equavalent to calculate the Kubo formular for 
the spin Hall conductivity\cite{SQHE-VMT-V}.  
The conductivity is represented by the Chern number and
quantized when the quasiparticle has an excitation gap 
as that in the $d$-wave vortex state\cite{SQHE-VMT-V}.
We have shown that the spin Hall conductivity is written by the Berry phase.
The spin transfer per a cycle per the boundary of the unit cell
is quantized and related to the Berry
  phase. The results remind us the adiabatic pumping, which is introduced by
Thouless with respect to the charge transport\cite{Thouless-pump}.
The result is also similar to the relation between
the spontaneous polarization and the Berry phase
in the crystalline dielectrics\cite{King-Smith-and-Vanderbilt-Resta}.
Essentially, the similarity comes from
the fact that the effects argued here
are caused by the closed adiabatic change
in the Bloch states with the finite energy gap.
With respect to the charge transport, a parallel discussion
have been made in the Bloch electron systems in the presence
of the electromagnetic field\cite{Goryo-Kohmoto2}.

As we mentioned before, the spin quantum Hall effect
in the vortex state of $d_{x^2-y^2}$-wave superconductor
has been pointed out\cite{SQHE-VMT-V}, but
it seems to have some difficulty to make a parallel discussion
in the {\it superconductors}.
Because of the Meissner effect,
it is not possible to have a magnetic field with a
finite homogeneous gradient which is essential to define the adiabatic
process on the closed loop in the parameter space.
The vortex states in $^3$He-A are suitable for our arguments because
$^3$He is the fermionic superfluid in which the spin current is
well defined, i.e. the spin rotation symmetry around $z$-axis is
  remained. In  contrast to the $d_{x^2-y^2}$-wave state,
the spin quantum Hall effect occurs spontaneously in $^3$He-A ,
i.e. one obtains a quantized spin Hall conductivity  
to calculate the Kubo formular in the absence of the vortices. The effect
comes from the broken time reversal symmetry and the broken parity in
the orbital part of the pairing symmetry\cite{SQHE-VY}.
But, the system does not have the finite spatial periodicity
and we can not make a parallel discussion also in this case.

We would like to comment on the fact that, in our argument,
the orbital part of the pairing symmetry is not crucial as long as
the quasiparticle spectrum in the vortex state has an excitation gap.
The spin part is crucial because the spin rotational symmetry
is needed to obtain well defined spin currents.

Several authors have made their efforts to find out a way
to measure spin transport\cite{spin-transport}.
Some experimental techniques to detect spin transfer
is highly desirable.

The authors are grateful to K. Maki, M. Sato,
Z. Te\v{s}anovi\'{c} and F. Zhou for useful discussions.

\end{multicols}

\end{document}